%% file: unitjobs_preemptive.tex
\documentclass[11pt]{article}

\usepackage{fullpage,latexsym,amsmath,amsfonts,epsfig}
\usepackage[latin1]{inputenc} 

\input{macros.tex}

\begin{document}


\title{Preemptive Scheduling of Equal-Length Jobs\\
to Maximize Weighted Throughput}

\author{
Philippe Baptiste
\thanks{CNRS LIX, Ecole Polytechnique, 91128 Palaiseau, France. baptiste@lix.polytechnique.fr.}
\and
Marek Chrobak%
\thanks{Department of Computer Science,
University of California,
Riverside, CA 92521.
Supported by NSF grants CCR-9988360 and CCR-0208856.
\{marek,wojtek\}@cs.ucr.edu.}
\and
Christoph Dürr%
\thanks{Laboratoire de Recherche en Informatique,
Université Paris-Sud,
91405 Orsay, France. durr@lri.fr.
Supported by the EU 5th framework programs QAIP
IST-1999-11234 and RAND-APX IST-1999-14036, and by
CNRS/STIC 01N80/0502 and 01N80/0607 grants.}
\and
Wojciech Jawor\footnotemark[2]
\and
Nodari Vakhania%
\thanks{Facultad de Ciencias,
Universidad Autonoma del Estado de Morelos,
62251 Cuernavaca, Morelos, Mexico.
nodari@servm.fc.uaem.mx.
Supported by CONACyT-NSF cooperative research grant E120.19.14. }
}

\maketitle

\begin{abstract}
We study the problem of computing a preemptive schedule of equal-length
jobs with given release times, deadlines and weights. Our goal is to maximize the
\emph{weighted throughput}, which
is the total weight of completed jobs. In Graham's notation this
problem is described as $(1|r_j; p_j\myeq p; \text{pmtn}|\sum w_j U_j)$.
We provide an $O(n^4)$-time algorithm for this problem, improving the
previous bound of $O(n^{10})$ by Baptiste~\cite{baptiste99a}.
\end{abstract}


\section{Introduction}\label{sec: introduction}

We study the following scheduling problem. We are given a set of $n$
jobs of the same integer length $p\ge 1$. For each job $j$ we are
also given three integer values: its release time $r_j$, deadline
$d_j$ and weight $w_j\ge 0$. Our goal is to compute a preemptive
schedule that maximizes the \emph{weighted throughput}, which is the
total weight of completed jobs. Alternatively, this is sometimes
formulated as minimizing the weighted number of late jobs. In Graham's
notation, this scheduling problem is described as $(1|r_j; p_j \myeq p;
\text{pmtn}|\sum w_j U_j)$, where $U_j$ is a 0-1 variable indicating
whether $j$ is completed or not in the schedule.

Most of the literature on job scheduling focuses on minimizing makespan,
lateness, tardiness, or other objective functions that depend on the
completion time of all jobs. 
Our work
is motivated by applications in real-time overloaded systems, where the
total workload often exceeds the capacity of the processor, and where
the job deadlines are critical, in the sense that the jobs that are not
completed by the deadline bring no benefit and may as well be removed
from the schedule altogether. In such systems, a reasonable goal is to
maximize the throughput, that is, the number of executed tasks. In more
general situations, some jobs may be more important than other. This can
be modeled by assigning weights to the jobs and maximizing the weighted
throughput (see, for example, \cite{KorSha95}).

The above problem $(1|r_j; p_j \myeq p; \text{pmtn}|\sum w_j U_j)$ was
studied by Baptiste~\cite{baptiste99a}, who showed that it can be solved
in polynomial time. His algorithm runs in time $O(n^{10})$. In this
paper we improve his result by providing an $O(n^4)$-time algorithm for
this problem.


\begin{figure}[htbp]
{\small
\begin{center}

{\tabcolsep = 0.025in
\begin{tabular}{llll} \hline
\parbox[t]{4cm}{\vspace{0.01in}
    $1|p_i\myeq p;r_j;\text{pmtn}|\sum U_j$\\
	$O(n\log n)$ \cite{lawler94}
\vspace{0.01in}
}
&
\parbox[t]{4cm}{\vspace{0.01in}
    $1|r_j;\text{pmtn}|\sum U_j$\\
	$O(n^5)$ \cite{lawler90}\\
	$O(n^4)$ \cite{baptiste99b}\\
\\
    $2|r_j;\text{pmtn}|\sum U_j$\\
	   NP-hard \cite{du92}
   \vspace{0.01in}
   }
&
&
\parbox[t]{4cm}{\vspace{0.01in}
    $1|p_i\myeq p;r_j|\sum U_j$\\
	   $O(n^3\log n)$ \cite{carlier81}
   \vspace{0.01in}
   }
\\ \hline
\parbox[t]{4cm}{\vspace{0.01in}
    $1|p_i\myeq p;r_j;\text{pmtn}|\sum w_j U_j$\\
	$O(n^4)$ [this paper]\\
	was $O(n^{10})$ \cite{baptiste99a}
\vspace{6mm}
\\
if $r_i<r_j\Rightarrow w_i\geq w_j$
\\
$O(n\log n)$
	\cite{lawler94}
\vspace{0.01in}
}
&
\parbox[t]{4cm}{\vspace{0.01in}
    $1|r_j;\text{pmtn}|\sum w_j U_j$\\
	   NP-hard \cite{garey_johnson79}\\
	   pseudo-polynomial \\ \cite{lawler90}
   \vspace{0.01in}
   }
&
\parbox[t]{4cm}{\vspace{0.01in}
	$P|p_i\myeq p;\text{pmtn}|\sum w_j U_j$\\
	       NP-hard \cite{brucker99} \\
	\\
    $Pm|p_i\myeq p;\text{pmtn}|\sum w_j U_j$\\
	       $O(n^{2^mm!})$ \cite{baptiste00b}\\
	       $O(nm(\max d_j)^m)$ \cite{baptiste00b}
       \vspace{0.01in}
	}
&
\parbox[t]{4cm}{\vspace{0.01in}
	$1|p_i\myeq p;r_j|\sum w_j U_j$\\
	   $O(n^7)$ \cite{baptiste99a} \\ \\
    $Pm|p_i\,=\,p;r_j|\sum w_j U_j$\\
	   $O(n^{6m+1})$ \cite{baptiste02b}
   \vspace{0.01in}
   }
\\ \hline
\end{tabular}
}

\caption{Complexity of some related throughput maximization
problems.}\label{fig:complexity}
\end{center}
}
\end{figure}


Figure~\ref{fig:complexity} shows some complexity results for related
scheduling problems where the objective function is to maximize
throughput. A more extensive overview can be found at Brucker and
Knust's website~\cite{brucker_web}. (That website, however, only
categorizes problems as NP-complete, polynomial, pseudo-polynomial or
open, without describing their exact time complexity.)


\section{Preliminaries}
\label{sec: preliminaries}


\paragraph{Terminology and notation.} We assume that the jobs on input
are numbered $1,2,\dots,n$. All jobs have the same integer length $p\ge
1$. Each job $j$ is specified by a triple $(r_j,d_j,w_j)$ of integers,
where $r_j$ is the release time, $d_j$ is the deadline, and $w_j\ge 0$
is the weight of $j$. Without loss of generality, we assume that $d_j\ge
r_j+p$ for all $j$ and that $\min_j r_j = 0$.

Throughout the paper, by a \emph{time unit $t$} we mean a time interval
$[t,t+1)$, where $t$ is an integer. A \emph{preemptive schedule} (or,
simply, a \emph{schedule}) $S$ is a function that assigns to each job
$j$ a set $S(j)$ of time units when $j$ is executed. Here, the term
``preemption" refers to the fact that the time units in $S$ may not be
consecutive.
We require that $S$ satisfies the following two conditions:
\begin{description}
\item{(sch1)} $S(j)\subseteq [r_j,d_j)$ for each $j$
(jobs are executed between their release times
	and deadlines.)
\item{(sch2)} $S(i)\cap S(j) = \emptyset$ for $i\neq j$
(at most one job is executed at a time.)
\end{description}
If $t\in S(j)$ then we say that (a unit of) $j$ is scheduled or executed
at time unit $t$. If $|S(j)| = p$, then we say that $S$
\emph{completes} $j$. The completion time of $j$ is $C_j = 1+ \max
S(j)$. Without loss of generality, we will be assuming that each job
$j$ is either completed ($|S(j)|=p$) or not executed at all ($S(j) =
\emptyset$).

The \emph{throughput} of $S$ is the total weight
of jobs that are completed in $S$, that is
$w(S) = \sum_{|S(j)|=p} w_j$.
Our goal is to find a schedule of all jobs with maximum throughput.

For a set of jobs $\calJ$, by $w(\calJ) = \sum_{j\in\calJ} w_j$ we
denote the total weight of $\calJ$. Given a set of jobs $\calJ$, if
there is a schedule $S$ that completes all jobs in $\calJ$, then we say
that $\calJ$ is \emph{feasible}. The restriction of $S$ to $\calJ$ is
called a \emph{full schedule} of $\calJ$.


\paragraph{Earliest-deadline schedules.} For two jobs $j,k$, we say that
$j$ is \emph{more urgent} than $k$ if $d_j < d_k$. It is well-known
that if $\calJ$ is feasible, then $\calJ$ can be fully scheduled using
the following \emph{earliest-deadline rule}: at every time step $t$,
execute the most urgent job among the jobs that have been released by
time $t$ but not yet completed. Ties can can be broken arbitrarily,
but consistently, for example, always in favor of lower numbered
jobs. If $S$ is any schedule (of all jobs), then we say that $S$ is
earliest-deadline if its restriction to the set of executed jobs is
earliest-deadline.

Since any feasible set of jobs $\calJ$ can be fully scheduled in time
$O(n\log n)$ using the earliest-deadline rule, the problem of computing
a schedule of maximum throughput is essentially equivalent to computing
a maximum-weight feasible set.


\begin{figure}[htbp]
\begin{center}
\epsfig{file=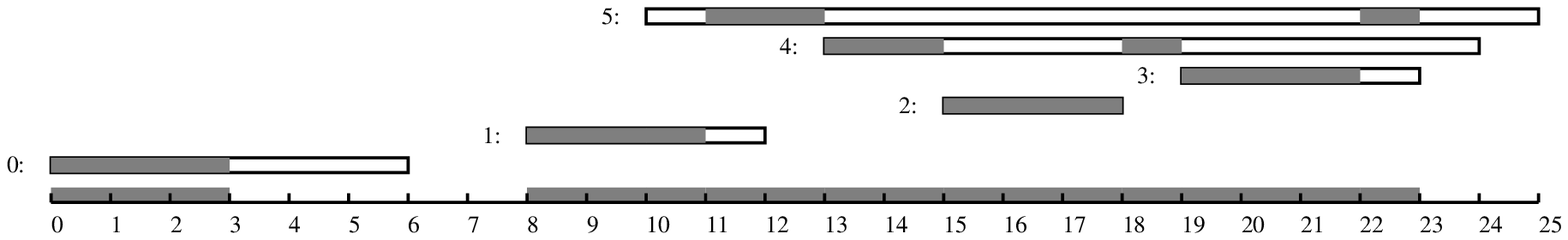,width=16cm}
\\
\epsfig{file=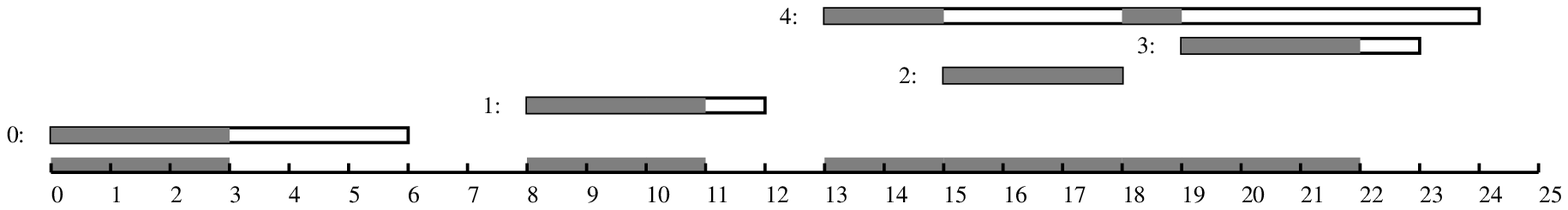,width=16cm}
\caption{Examples of earliest-deadline schedules with $p=3$.
The rectangles represent intervals $[r_j,d_j)$, and the
shaded areas show time units where jobs are executed.
The first  schedule consists of two distinct blocks.
After removing the least urgent job $5$,
the second block splits into several smaller blocks.}
\label{fig:edd}
\end{center}
\end{figure}


Each earliest-deadline schedule $S$ has the following structure. The
time axis is divided into busy intervals (when jobs are being executed)
called \emph{blocks} and idle intervals called \emph{gaps}. Each block
is an interval $[r_i,C_j)$ between a release time $r_i$ and a completion
time $C_j$, and it satisfies the following two properties: (b1) all
jobs executed in this block are not released before $r_i$, and (b2)
$C_j$ is the first completion time after $r_i$ such that all jobs in $S$
released before $C_j$ are completed at or before $C_j$. Note that $C_j -
r_i = ap$, for $a$ equal to the number of jobs executed in this block.
Figure~\ref{fig:edd} shows two examples of earliest-deadline schedules.

In some degenerate situations, where the differences between release
times are multiples of $p$,  a gap can be empty, and
the end $C_j$ of one block then equals  the beginning $r_m$ of the next
block.

The above structure is recursive, in the following sense. Let $k$ be the
least urgent job scheduled in a given block
$[r_i,C_j)$. Then the last completed job is $k$. Also, when we remove
job $k$ from the schedule, without any further modifications, we obtain
again an earliest-deadline-schedule for the set of remaining jobs (See
Figure~\ref{fig:edd}). The interval $[r_i,C_j)$ may now contain several
blocks of this new schedule.


\subsection{An $O(n^4)$-Time Algorithm}
\label{sec: weighted case}

We assume that the jobs are ordered $1,2,\dots,n$ according to
non-decreasing deadlines, that is $d_1 \le d_2 \le \ldots \le d_n$.
Without loss of generality we may assume that job $n$ is a ``dummy" job
with $w_n = 0$ and $r_n = d_{n-1}$ (otherwise, we can add one such
additional job). We use letters $i,j,k,l \in[1,n]$ for job identifiers,
and $a,b \in [0,n]$ for numbers of jobs.

Given an interval $[x,y)$,
define a set $\calJ$ of jobs to be $(k,x,y)$\emph{-feasible} if
\begin{description}
\item{(f1)} $\calJ\subseteq \braced{1,2,\dots,k}$,
\item{(f2)} $r_j\in[x,y)$ for all $j\in\calJ$, and
\item{(f3)} $\calJ$ has a full schedule in $[x,y)$ (that is,
all jobs are completed by time $y$.)
\end{description}
An earliest-deadline schedule of a $(k,x,y)$-feasible set of jobs will
be called a $(k,x,y)$\emph{-schedule}. If ties are broken consistently,
then there is a 1-1 correspondence between feasible sets of jobs and
their earliest-deadline schedules. Thus, for the sake of simplicity, we
will use the same notation $\calJ$ for a feasible set of jobs and for
its earliest-deadline schedule.

Note that if $e$ is the job with the earliest release time, then
an optimal $(n,r_e,r_n)$-schedule is also an optimal schedule to
the whole instance. The idea of the algorithm is to compute optimal
$(k,r_i,r_j)$-schedules $\calF^k_{i,j}$ in bottom-up order, using
dynamic programming. As there does not seem to be an efficient way to
express $\calF^k_{i,j}$ in terms of such sets for smaller instances,
we use two auxiliary optimal schedules denoted $\calG^k_{i,a}$ 
and
 $\calH^k_{i,j}$ on which we impose some additional restrictions.

We first define the values $F^k_{i,j}$, $G^k_{i,j}$, and $H^k_{i,a}$
that are meant to represent the weights of the corresponding schedules
mentioned above. The interpretation of these values is as follows:

\medskip

\begin{tabular}{lcp{5in}}
$F^k_{i,j}$ & = & the optimal weight of a $(k,r_i,r_j)$-schedule.
		\\
		\\
$G^k_{i,a}$ &=& the optimal weight of a $(k,r_i,r_i + ap)$-schedule that
consists of a single block starting at time $r_i$ and ending at $r_i + ap$.
		\\
		\\
$H^k_{i,j}$ & = & the optimal weight of a $(k,r_i,r_j)$-schedule
that has no gap between $r_i$ and $r_{k+1}$.

\end{tabular}

In $F^k_{i,j}$ and $H^k_{i,j}$ we assume that $r_i \le r_j$.
In $H^k_{i,j}$ we additionally assume that $k<n$ and
$r_i \le r_{k+1}$.

\medskip

We now give recursive formulas these values. In these formulas we use
the following auxiliary functions:
\begin{eqnarray*}
\nofjobs(x,y) &=& \min \braced{ n, \ceiling { \frac{y-x}{p}} -1 } 
	\\
\earliestjob(x)   &=& \argmin_i \braced{r_i : r_i \ge x}
        \\
\nextjob(x)   &=& \argmin_i \braced{r_i : r_i > x}
\end{eqnarray*}
Thus $\nofjobs(x,y)$ is the maximum number of jobs (but not more
than $n$) that can be executed between $x$ and $y$ (ignoring
release times and deadlines), such that the interval $[x,y)$ is not
completely filled. For $x\le r_n$, $\earliestjob(x)$ denotes
the first job released at or after $x$. Similarly, for $x < r_n$,
$\nextjob(x)$ is the first job released strictly after $x$.
(Ties can be broken arbitrarily).

\medskip
\noindent 
{\it Values $F^k_{i,j}$.\ } 
If $r_j = r_i$ then $F^k_{i,j} = 0$. Otherwise, $F^k_{i,j}$ is defined
inductively as follows:
\begin{eqnarray*}
F^k_{i,j}
&=& \max\left\{\begin{array}{lclr}
\displaystyle
F^{k}_{\nextjob(r_i),j}
&&
& (F1)
\\ \\
\displaystyle
\max_{\substack{1\le a \le n \\r_i+ap \le r_j} }
\braced{G^k_{i,a} + F^k_{\earliestjob(r_i+ap),j} }
&&
& (F2)
\end{array} \right.
\end{eqnarray*}
Note that in (F1) $\nextjob(r_i)$ is well defined since $r_i<r_j$, and in
(F2) $\earliestjob(r_i+ap)$ is well defined since $r_i+ap\le r_j$.

\medskip
\noindent
{\it Values $G^k_{i,a}$.\ } 
If $k=0$ or $a=0$, then $G^k_{i,a} = 0$. If $r_k\not\in[r_i,r_i+(a-1)p]$
or $d_k<r_i+ap$
then $G^k_{i,a}=G^{k-1}_{i,a}$.
Otherwise, $G^{k}_{i,a}$ is defined as follows:
\begin{eqnarray*}
G^{k}_{i,a}
&=& \max\left\{\begin{array}{lclr}
\displaystyle
G^{k-1}_{i,a}
&&
& (G1)
\\ \\
\displaystyle
G^{k-1}_{i,a-1}+w_k
&&
& (G2)
\\ \\
\displaystyle
\max_{r_k < r_l < r_i+ap}
                \braced{H^{k-1}_{i,l}+G^{k-1}_{l,\nofjobs(r_l,r_i+ap)} +w_k}
&&
& (G3)
\end{array} \right.
\end{eqnarray*}

\medskip
\noindent
{\it Values $H^k_{i,j}$.\ } 
If $r_j = r_i$ then $H^k_{i,j}=0$. If $k=n$ or
$r_{k+1}\not\in[r_i,r_j]$ then $H^k_{i,j}$ is undefined. For other
values $H^{k}_{i,j}$ is defined inductively as follows:
\begin{eqnarray*}
H^{k}_{i,j}
&=& \begin{array}{lclr}
\displaystyle
\max_{\substack{0\le a \le n\\
	r_{k+1} \le r_i+ap \le r_j}}
\braced{G^k_{i,a} + F^k_{\earliestjob(r_i+ap),j} }
&&
& (H)
\end{array}
\end{eqnarray*}
%


\begin{figure}[htbp]

\smallskip
\noindent
\epsfig{file=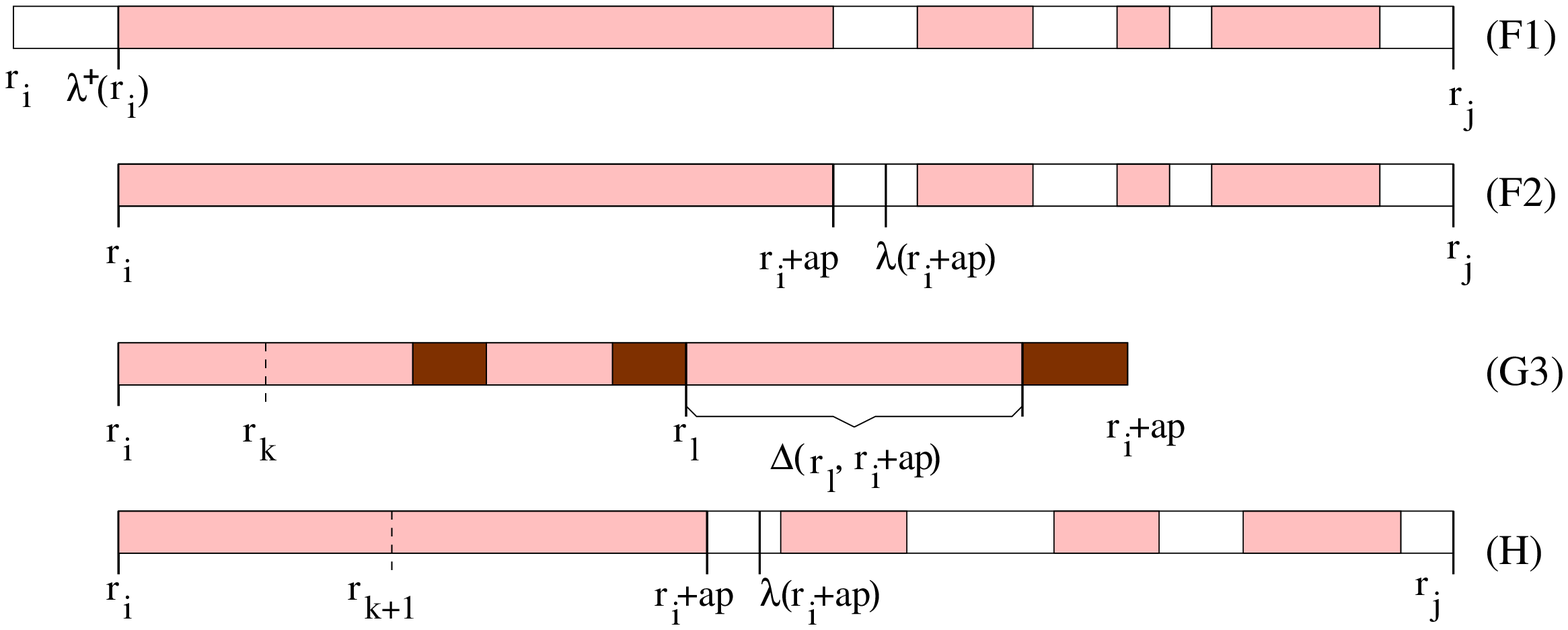,width=16cm} \hfill

\caption{Graphical explanation of the recursive formulas for
$F^{k}_{i,j}$, $G^{k}_{i,a}$ and $H^{k}_{i,j}$. Shaded regions
show blocks. In (G3), the whole schedule is one block, and
darker shade shows where $k$ is executed.}
\label{fig:computing fgh}
\end{figure}


\paragraph{Algorithm~{\DP}.}
The algorithm first computes the values $F^k_{i,j}, G^{k}_{i,a},
H^k_{i,j}$ bottom-up. The general structure of this first stage
is as follows:

\begin{center}
\begin{minipage}{3in}
\begin{tabbing}
\hskip 0.3in\= \hskip 0.3in\= \hskip 0.3in\= \hskip 0.3in\= \kill \\
\ttfor $k\assign 0$ \ttto $n$ \ttdo \\
\> \ttfor $i \assign n$ \ttdownto $1$ \ttdo \\
\>\> \ttfor $a \assign 0$ \ttto $n$ \ttdo \\
\>\>\>    compute $G^k_{i,a}$  \\
\>\> \ttfor $j \assign i$ \ttto $n$ \ttdo \\
\>\>\>    compute $F^k_{i,j}$ and $H^k_{i,j}$ 
\end{tabbing}
\end{minipage}
\end{center}

The values $F^k_{i,j}$, $G^k_{i,a}$, and $H^k_{i,j}$ are computed
according to their recursive definitions, as given earlier. At each
step, we record which value realized the maximum.

\smallskip

In the second stage, we construct an optimal
schedule $\calF^n_{e,n}$, where $e$ is the job with earliest deadline.
This is achieved by starting with $\calF^n_{e,n}$ and recursively
reconstructing optimal schedules $\calF^k_{i,j}$, $\calG^k_{i,a}$
and $\calH^k_{i,j}$ that realize weights $F^k_{i,j},
G^k_{i,a}$, and $H^k_{i,j}$, respectively, according to the
following procedure.

\smallskip
\noindent
{\it Computing $\calF^k_{i,j}$.}
If $F^k_{i,j}=0$, return $\calF^k_{i,j} = \emptyset$. If
$F^k_{i,j}$ was maximized by choice (F1), let $\calF^k_{i,j} =
\calF^{k-1}_{\nextjob(r_i),j}$. If $F^k_{i,j}$ was maximized
by choice (F2), let $\calF^k_{i,j} = \calG^k_{i,a} \cup
\calF^k_{\earliestjob(r_i+ap),j}$, where $a$ is the integer that realizes
the maximum.

\smallskip
\noindent
{\it Computing $G^k_{i,a}$.}
If $G^k_{i,a}=0$, return $\calG^k_{i,a} = \emptyset$. If $G^k_{i,a}$
is realized by choice (G1), let $\calG^k_{i,a} =\calG^{k-1}_{i,a}$.
If $G^k_{i,a}$ is realized by choice (G2),
let $\calG^k_{i,a}= \calG^{k-1}_{i,a-1}\cup \braced{k}$.
If $G^k_{i,a}$ is realized by choice (G3), let $\calG^k_{i,a}=
\calH^{k-1}_{i,l} \cup \calG^{k-1}_{l,\nofjobs(r_l,r_i+ap)}
\cup\braced{k}$, where $l$ is the job that realizes the maximum in (G3).

\smallskip
\noindent
{\it Computing $\calH^k_{i,j}$.}
If $H^k_{i,j}=0$, return $\calH^k_{i,j} = \emptyset$. Otherwise,
$\calH^k_{i,j} = \calG^k_{i,a} \cup \calF^k_{\earliestjob(r_i+ap),j}$,
where $a$ is the integer that realizes the maximum.


\bigskip

\begin{theorem}\label{thm: w is maximum weight}
Algorithm~{\DP} correctly computes a maximum-weight feasible set
of jobs and it runs in time $O(n^4)$.
\end{theorem}

\begin{proof}
The time complexity is quite obvious: We have $O(n^3)$ values
$F^k_{i,j}$, $G^k_{i,a}$ , $H^k_{i,j}$, and they can be stored
in 3-dimensional tables. The functions $\nofjobs(\cdot,\cdot)$,
$\nextjob(\cdot)$, and $\earliestjob(\cdot)$ can be precomputed.
Then each entry in these tables can be computed in time $O(n)$. The
reconstruction of the schedules in the second part takes only time
$O(n)$.

\smallskip

To show correctness, we need to prove two claims:
\begin{description}
\item{Claim~1:}
\begin{description}
\item{(1f)}
$w(\calF^k_{i,j})= F^k_{i,j}$ and
$\calF^k_{i,j}$ is a $(k,r_i,r_j)$-schedule.
\item{(1g)}
$w(\calG^k_{i,a})= G^k_{i,a}$ and
$\calG^k_{i,a}$ is $(k,r_i,r_i+ap)$-schedule that consists of a 
single block of $a$ jobs starting at $r_i$.
\item{(1h)}
$w(\calH^k_{i,j})= H^k_{i,j}$ and
$\calH^k_{i,j}$ is $(k,r_i,r_j)$-schedule that has no
gap before $r_{k+1}$ (assuming that $k<n$ and $r_i \le r_{k+1} \le r_j$.)
\end{description}
\item{Claim~2:}
\begin{description}
\item{(2f)}
If $\calJ$ is a $(k,r_i,r_j)$-schedule
then $w(\calJ) \le F^k_{i,j}$.
\item{(2g)}
If $\calJ$ is a $(k,r_i,r_i+ap)$-schedule that is
a single block of $a$ jobs
then $w(\calJ) \le G^k_{i,a}$.
\item{(2h)}
If $\calJ$ is a $(k,r_i,r_j)$-schedule that has no
gap before $r_{k+1}$ then $w(\calJ) \le H^k_{i,j}$
(assuming that $k<n$ and $r_i \le r_{k+1} \le r_j$.)
\end{description}
\end{description}

We prove both claims by induction. We first define a partial order on all
function instances $F^k_{i,j}$, $G^k_{i,a}$ and $H^k_{i,j}$. We first
order them in order of increasing $k$. For a fixed $k$, 
we order them in order of increasing length of their time
intervals, that is, $r_j - r_i$ for $F^k_{i,j}$ and $H^k_{i,j}$,
and $ap$ for $G^k_{i,a}$. Finally, for a fixed $k$, $i$ and $j$,
we assume that $F^k_{i,j}$ is before $H^k_{i,j}$.
The induction will proceed with respect to this ordering.

We now prove Claim~1. The base cases are when $k=0$ or $a=0$ in $G^k_{i,a}$,
or $r_i =r_j$ in $F^k_{i,j}$ and $H^k_{i,j}$. In all these cases Claim~1
holds trivially. We now examine the inductive steps.

To prove (1f), if $\calF^k_{i,j}$ was constructed from case (F1), the
claim holds by induction. If $\calF^k_{i,j}$ was constructed from
case (F2), let $a$ be the integer that realizes the maximum and $l =
\earliestjob(r_i+ap)$. Since $r_i+ap \le r_l$, sets $\calG^k_{i,a}$ and
$\calF^k_{l,j}$ are disjoint, and so are the intervals $[r_i,r_i+ ap)$,
$[r_l,r_j)$. Thus both $\calG^k_{i,a}$ and $\calF^k_{l,j}$ can be fully
scheduled in $[r_i,r_j)$ and $w(\calF^k_{i,j}) = w(\calG^k_{i,a}) +
w(\calF^k_{l,j}) = G^k_{i,a} + F^k_{l,j} = F^k_{i,j}$, by induction.

To prove (1g), if $G^k_{i,a}$ is realized by case (G1), the claim is
obvious.
In case (G2), we have $k\notin\calG^{k-1}_{i,a-1}$, 
$r_k \le r_i+(a-1)p$, and $d_k \ge r_i+ap$. Thus we can 
schedule $\calG^{k-1}_{i,a-1}$, and then schedule $k$ at $r_i + (a-1)p$.
By induction, $w(\calG^k_{i,a}) = w(\calG^{k-1}_{i,a-1}) + w_k
= G^{k-1}_{i,a-1} + w_k =  G^k_{i,a}$.
In case (G3), let $l$ be the job that realizes the maximum
and $b = \nofjobs(r_l,r_i+ap)$. The sets $\calH^{k-1}_{i,l}$ and
$\calG^{k-1}_{l,b}$ are disjoint and so are the intervals $[r_i,r_l)$,
$[r_l, r_l+bp)$. By the definition of $b$, we have
$r_l + bp < r_i + ap$, so there is a non-zero idle time in $\calH^{k-1}_{i,l}\cup
\calG^{k-1}_{l,b}$ between $r_l+bp$ and $r_i+ap$. Since the total
interval $[r_i, r_i+ap)$ has length $ap$, the total idle time 
in this interval must be at least $p$. Moreover, all gaps occur
after $r_k$. This implies that we can schedule job $k$
in these idle intervals. Also, note that
$w(\calG^k_{i,a})= w(\calH^{k-1}_{i,l}) + w(\calG^{k-1}_{l,b}) +w_k =
H^{k-1}_{i,l} + G^{k-1}_{l,b} +w_k = G^k_{i,a}$, so the claim holds.

To prove (1h), let $a$ be the integer that realizes the maximum in (H) and
$l = \earliestjob(r_i+ap)$. As before, since $r_i+ap \le r_l$, sets
$\calG^k_{i,a}$ and $\calF^k_{l,j}$ are disjoint, and so are the
intervals $[r_i,r_i+ ap)$, $[r_l,r_j)$. Thus both $\calG^k_{i,a}$
and $\calF^k_{l,j}$ can be fully scheduled in $[r_i,r_j)$ and
$w(\calH^k_{i,j}) = w(\calG^k_{i,a}) + w(\calF^k_{l,j}) = G^k_{i,a} +
F^k_{l,j} = H^k_{i,j}$, by induction.

\bigskip

We now show Claim~2. Again, we proceed by induction with respect to
the ordering of the instances described before the proof of Claim~1.
The claim holds trivially for the base cases. We now consider the
inductive step.

To prove (2f), we have two cases. If $\calJ$ does not start at $r_i$,
then it cannot start earlier than at $r_m$, for $m = {\nextjob(r_i)}$,
so the claim follows by induction. If $\calJ$ starts at $r_i$, let
$a$ be the length of its first block. The second block (if any) cannot start earlier
than at $r_l$, for $l=\earliestjob(r_i+ap)$. (Note that there might be no gap between the blocks.)
We partition $\calJ$ into two sets:
$\calJ_1$ containing the jobs scheduled in $[r_i, r_i+ap)$ as a single
block, and $\calJ_2$ containing the jobs scheduled in $[r_l, r_j)$. By
induction, $w(\calJ) = w(\calJ_1) + w(\calJ_2) \le G^k_{i,a} + F^k_{l,j}
\le F^k_{i,j}$.

We now prove (2g). If $k\notin \calJ$ then $\calJ$ is a
$(k-1,r_i,r_i+ap)$-schedule, so $w(\calJ) \le G^{k-1}_{i,a} \le
G^k_{i,a}$, by induction and by case (G1). Now assume that $k\in\calJ$.
If job $k$ has not been interrupted, then $\calJ - \braced{k}$ is
a $(k-1,r_i, r_i+(a-1)p)$-schedule. Thus, by induction and (G2), $w(\calJ) \le
G^{k-1}_{i,a-1} + w_{k} \le G^k_{i,a}$.

Otherwise, let $l$ be the last job that interrupted $k$. Starting at
$r_l$, $\calJ$ executes $b = \nofjobs(r_l,r_i+ap)$ jobs with deadlines
smaller than $d_k$, after which it executes a portion $r_i+ap -
r_l - bp > 0$ of job $k$. We partition $\calJ - \braced{k}$ into
two sets: $\calJ_1$ containing the jobs scheduled before $r_l$, and
$\calJ_2$ containing the jobs scheduled after $r_l$. Note that $\calJ_1
\cup \calJ_2 = \calJ - \braced{k}$, since the jobs scheduled before
$r_l$ must also be completed before $r_l$ and the other jobs cannot
be released yet. By induction, sets $\calJ_1$ is a $(k-1,r_i,r_l)$-schedule
in which the first block starts at $r_i$ and ends after
$r_k$, and $\calJ_2$ is a single block starting at $r_l$ and ending at
$r_l+bp$. Thus, by induction and (G3), $w(\calJ) = w(\calJ_1)+w(\calJ_2) + w_k \le
H^{k-1}_{i,l} + G^{k-1}_{l,b} + w_k \le G^k_{i,a}$.

The proof of (2h) is similar. We have two subcases.
Suppose first that $r_{k+1} = r_i$. By induction, we have
$w(\calJ) \le F^k_{i,j} = H^k_{i,j}$, since in this case
we can choose $a=0$ in (H). If $r_{k+1} > r_i$, then
the first block of $\calJ$ starts at $r_i$ and ends after $r_{k+1}$. Let $a$ be the
length of its first block. The second block (if any) cannot start
earlier than at $r_l$, for $l= \earliestjob(r_i+ap)$. We partition $\calJ$ into two sets
$\calJ_1$ containing the jobs scheduled in $[r_i, r_i+ap)$ as a single
block and $\calJ_2$ containing the jobs scheduled in $[r_l, r_j)$. By
induction, $w(\calJ) = w(\calJ_1) + w(\calJ_2) \le G^k_{i,a} + F^k_{l,j}
\le H^k_{i,j}$.
\hfill\end{proof}


\section{Final Remarks}

Several open problems remain. Although our running time $O(n^4)$ for the
scheduling problem $(1|r_j; p_j\myeq p; \text{pmtn}|\sum w_j U_j)$ is
substantially better than the previous bound of $O(n^{10})$, it would
be interesting to see whether it can be improved further. Similarly, it
would be interesting to improve the running time for the non-preemptive
version of this problem, $(1|r_j; p_j\myeq p|\sum w_j U_j)$, which is
currently $O(n^7)$ \cite{baptiste99b}.

In the multi-processor case, the weighted version is known to be
NP-complete \cite{brucker99}, but the non-weighted version remains
open. More specifically, it is not known whether the problem $(P|r_j;
p_j\myeq p; \text{pmtn}|\sum U_j)$ can be solved in polynomial time.
(One difficulty that arises for $2$ or more processors is that we cannot
restrict ourselves to earliest-deadline schedules. For example, an
instance consisting of three jobs with feasible intervals $(0,3)$,
$(0,4)$, and $(0,5)$ and processing time $p=3$ is feasible, but the
earliest-deadline schedule will complete only jobs $1$ and $2$.) In the
multi-processor case, one can also consider a preemptive version where
jobs are not allowed to migrate between processors.


\bibliographystyle{alpha} \bibliography{scheduling}

\end{document}

%% file: macros.tex
\newcommand{\calF}{{\cal F}}

\newcommand{\calG}{{\cal G}}
\newcommand{\calH}{{\cal H}}

\newcommand{\calJ}{{\cal J}}

\newcommand{\braced}[1]{{ \left\{ #1 \right\} }}

\newcommand{\ceiling}[1]{{ \left\lceil #1 \right\rceil }}

\newcommand{\nofjobs}{{\Delta}}
\newcommand{\earliestjob}{{\lambda}}
\newcommand{\nextjob}{{\lambda^+}}

\newcommand{\myeq}{{=}}

\newcommand{\DP}{{\mbox{\sc DP}}}

\newcommand{\assign}{{\,\leftarrow\,}}

\newcommand{\ttfor}{ {\mbox{\bf for}} }
\newcommand{\ttto}{ {\mbox{\bf to}} }
\newcommand{\ttdownto}{ {\mbox{\bf downto}} }
\newcommand{\ttdo}{ {\mbox{\bf do}} }

\newcommand{\argmin}{{\mbox{\rm argmin}}}

\newtheorem{theorem}{Theorem}[section]

\newenvironment{proof}{{\it Proof:\/}}{$\Box$\vskip 0.1in}
                {$\spadesuit$\smallskip}

\newenvironment{bigeqn*}{\large\begin{eqnarray*}}{\end{eqnarray*}}
